\magnification=1400

\centerline{\bf STRANGENESS VIOLATING DIBARYON DECAY}
\medskip
\centerline{ Sheldon L. Glashow }
\smallskip\centerline{\it  Department of Physics, Boston
University, Boston, MA 02215}

\bigskip
\centerline{\it \bf Abstract}

Non-standard physics may induce detectable
flavor-changing $\Delta B=2$ interactions without inducing their
flavor-conserving counterparts. Searches for $n$-$\overline n$
oscillations do not constrain such interactions, thereby motivating
dedicated searches for $\Delta B=2$ nuclear decays into
strange final states. In particular, the simple model herein proposed
enables dibaryon decay exclusively into states with   $S=2$.  \bigskip

We  begin with a brief parable. The flavor-conserving neutral currents
of the electroweak theory  were discovered  in 1973,
but  could (and should) have been found a decade earlier. As
Tini Veltman wrote,${}^1$  ``There was [in the 1960s]  a rather heavy bias
against [neutral current phenomena]. In quite different circumstances
they  had been found to be absent with a high
degree of certainty... Experimentally, it had been made sure that even
if there were such events they would not be discovered... There was
not much interest in that question at that time, something that
changed after 1971'' when the electroweak
theory was proven to be sensible.  In other words, the failure to observe
flavor-changing neutral currents  in strange particle decays
diverted experimenters from performing a timely search for flavor-conserving 
neutral currents in neutrino physics.  
\medskip

The situation regarding $\Delta B=2$ processes appears to be inversely
analogous. Novel interactions at a large mass scale could lead to  two
distinct observable phenomena: neutron-antineutron oscillations and
dibaryon decays. Both processes have been searched for, but as yet
neither has been seen.${}^2$ Far more sensitive proposed searches for
$n$-$\overline{n}$ oscillations${}^3$ may have dampened
 enthusiasm for further searches for  dibaryon decay.  However,
one easily may build models that generate only strangeness-changing
$\Delta B=2$ operators. These  do not generate strangeness-conserving
$\Delta B=2$ operators,  nor do they yield $n$-$\overline{n}$
oscillations.\footnote\dag{Has anyone looked for strangeness-changing
 dibaryon decays?
The PDG offers constraints on 13 $\Delta B=2$ decay modes, nary
a one involving kaons.${}^4$} \medskip

 An example of such a model (by no means unique) 
 serves as an existence proof.
 We introduce  several massive
spinless color triplets, each  coupled to a pair of
 right-handed quarks.
 Color antisymmetry of these quark pairs demands
that they also be   flavor antisymmetric. The three relevant bosons,
$R_{ud}$, $R_{us}$, 
and $R_{ds}$, 
 possess
the  trilinear
self coupling
 $\epsilon_{ijk}(R_{ud}^i R_{us}^j R_{ds}^k)$, which induces 
the effective
$\Delta B=2$ six-fermion interaction:
$$M^{-5}\,\epsilon^{ijk}
(ud)_i(us)_j\,(ds)_k$$
 where $M$ is the relevant mass scale.  
This interaction mediates the $\Delta B=2$ decay of a pair of
nucleons,
but exclusively
into states with $S=2$,  {\it e.g.,} a pair of kaons.
Note that flavor mixing
of the quark masses cannot yield the flavor-conserving $\Delta B=2$
term $uudddd$ and thus cannot induce $n$-$\overline n$ oscillations.

 \medskip

A one-loop diagram (with flavor mixing between $R_{ud}$ and $R_{us}$)
induces the $\Delta S=2$ four-fermion interaction
$M^{-2}ss\overline{d}\overline{d}$.  For this operator to have a
measurable effect on the neutral kaon system, $M$ would have to lie
below $\sim\! 100\;$TeV, whence strangeness-changing  dibaryon decay
would proceed so rapidly that it could not have escaped detection.
Conversely, dedicated searches for $\Delta S=2$ dibaryon decay can be
sensitive to values of $M$ beyond $ 1000\;$TeV.  Thus we see the need
for experimenters at Super-K and elsewhere to pursue the search for
dibaryon decays into strange final states, whether $S=2$ as in our
model, or $S=1\; {\rm or\; 3}$ in other conceivable models.  The
discovery of such phenomena would open a window into new physics 
 lying  beyond the reach of the LHC.  \medskip

\centerline{\bf Acknowledgement}\medskip

I am indebted to Paul H. Frampton for informing me of current searches
(or the lack thereof)
for dibaryon decay, and for challenging me to devise a scheme  which
yields such decays without inducing $n$-$\overline n$ oscillations. 
This research was supported in part by the Department of Energy under
contract number
DE-FG02-01ER-40676.
\vfill\eject
\centerline{\bf References}
\bigskip

\noindent
[1] M. Veltman in ``Facts and Mysteries in Particle Physics,''
(World Scientific, 2003, Singapore) p. 214.
\smallskip

\noindent [2] For an up-to-date review, see: R.N.  Mohapatra,
JPhys. G: Nucl.Phys.  36(2009)104006.
\smallskip

\noindent [3] {\it E.g.,} Y. Kamyshkov, {\tt  ArXiv:hep-ex/0211006.} 
\smallskip

\noindent [4] C. Amsler {\it et al.} (Particle Data Group), PL B667
(2008) and 2009 update for 2010 ed. (p 9 of proton listing).

\bye